\documentclass[10pt]{article}

\usepackage[margin=0.75in, top=1in, bottom=1in]{geometry}
\usepackage{graphicx}
\usepackage{hyperref}
\usepackage{url}
\usepackage{microtype}
\usepackage{parskip}
\usepackage{titlesec}
\usepackage{xcolor}
\usepackage{amsmath}
\usepackage{fontenc}
\usepackage{abstract}
\usepackage{caption}
\usepackage{datetime2}
\usepackage{setspace}
\usepackage{dblfnote}   
\usepackage{footmisc}   
\usepackage[normalem]{ulem}   

\newif\ifshowchanges
\showchangesfalse      
\ifshowchanges
  \newcommand{\new}[1]{\textcolor{red}{#1}}
  \newcommand{\del}[1]{\textcolor{blue}{\sout{#1}}}
  \newcommand{\newcite}[1]{{\hypersetup{citecolor=red}\cite{#1}}}
\else
  \newcommand{\new}[1]{#1}
  \newcommand{\del}[1]{}
  \newcommand{\newcite}[1]{\cite{#1}}
\fi

\titleformat{\section}{\normalsize\bfseries\uppercase}{\thesection}{0.5em}{}
\titleformat{\subsection}{\normalsize\bfseries}{\thesubsection}{0.5em}{}
\titleformat{\subsubsection}[runin]{\normalsize\bfseries}{}{0em}{}[.\enspace]
\titlespacing{\section}{0pt}{10pt}{4pt}
\titlespacing{\subsection}{0pt}{8pt}{2pt}
\titlespacing{\subsubsection}{0pt}{6pt}{0pt}

\hypersetup{
    colorlinks=true,
    linkcolor=blue,
    citecolor=blue,
    urlcolor=blue
}

\renewenvironment{abstract}{%
  \small\textbf{Abstract}---\ignorespaces
}{\par\vspace{6pt}}

\newcommand{\optin}[1]{}

\newcommand{\af}[1]{\textsuperscript{\scriptsize #1}}

\title{\large\textbf{NeuroAI and Beyond: Bridging Between Advances in
Neuroscience and Artificial Intelligence}}

\author{}   
\date{}

\begin{document}

\begin{onecolumn}
\maketitle
\vspace{-18pt}

\begin{center}
\small
Anthony Zador\af{28,\dag},
Jean-Marc Fellous\af{11,30,\dag},
Terrence Sejnowski\af{11,26,\dag},
Gina Adam\af{1},
James B Aimone\af{2},
Akwasi Akwaboah\af{3},
Yiannis Aloimonos\af{4},
Carmen Amo Alonso\af{5},
Chiara Bartolozzi\af{6},
Michael J.\ Bennington\af{7},
Michael Berry\af{8},
Bing W.\ Brunton\af{10},
Gert Cauwenberghs\af{11},
Hillel J.\ Chiel\af{12},
Tobi Delbruck\af{13},
John Doyle\af{14},
Jason Eshraghian\af{15},
Ralph Etienne-Cummings\af{3},
Cornelia Ferm\"{u}ller\af{4},
Matthew Jacobsen\af{4},
Christopher Kanan\af{17},
Ali A.\ Minai\af{19},
Barbara Oakley\af{20},
Alexander G.\ Ororbia II\af{21},
Joe Paton\af{22},
Blake Richards\af{23},
Yulia Sandamirskaya\af{24},
Abhronil Sengupta\af{25},
Shihab Shamma\af{4},
Michael P. Stryker\af{27},
Seong Jong Yoo\af{4},
Steven W.\ Zucker\af{29}
\end{center}
\vspace{2pt}

\noindent\rule{\linewidth}{0.4pt}
\begin{abstract}
Neuroscience and Artificial Intelligence (AI) have made impressive progress in recent years but remain only loosely interconnected. Based on a workshop convened by the National Science Foundation in August 2025, we identify three fundamental capability gaps in current AI—the limited ability to interact robustly with the physical world, inadequate learning that produces brittle systems, and unsustainable energy and data inefficiency—and describe the neuroscience principles that address each: co-design of body and controller, prediction through interaction, multi-scale learning with neuromodulatory control, hierarchical distributed architectures, and sparse event-driven computation. We present a research roadmap organized around these principles at near-, mid-, and long-term horizons. We argue that realizing this program requires a new generation of researchers trained across the boundary between neuroscience and engineering, and describe the institutional conditions—interdisciplinary training, hardware access, community standards, and ethics—needed to support them. We conclude that NeuroAI, neuroscience-informed artificial intelligence, has the potential to overcome limitations of current AI while deepening our understanding of biological neural computation.
\end{abstract}
\noindent\rule{\linewidth}{0.4pt}

\vspace{2pt}
{\tiny\setlength{\parskip}{0pt}\setlength{\parindent}{0pt}\setlength{\baselineskip}{7pt}\noindent
\textsuperscript{1}Dept.\ Electrical \& Computer Engineering, George Washington University, Washington, DC, USA.
\textsuperscript{2}Neural Exploration \& Research Laboratory, Sandia National Laboratories, Albuquerque, NM, USA.
\textsuperscript{3}Dept.\ Electrical \& Computer Engineering, Johns Hopkins University, Baltimore, MD, USA.
\textsuperscript{4}Institute for Advanced Computer Studies, University of Maryland, College Park, MD, USA.
\textsuperscript{5}Dept.\ Aeronautics \& Astronautics, Stanford University, Stanford, CA, USA.
\textsuperscript{6}Event-Driven Perception for Robotics, Italian Institute of Technology, Genoa, Italy.
\textsuperscript{7}Dept.\ Mechanical Engineering, Carnegie Mellon University, Pittsburgh, PA, USA.
\textsuperscript{8}Dept.\ Neuroscience, Princeton University, Princeton, NJ, USA.
\textsuperscript{9}Mila -- Quebec AI Institute \& Universit\'{e} de Montr\'{e}al, Montr\'{e}al, QC, Canada.
\textsuperscript{10}Dept.\ Biology, University of Washington, Seattle, WA, USA.
\textsuperscript{11}Institute for Neural Computation, UC San Diego, La Jolla, CA, USA.
\textsuperscript{12}Depts.\ Biology, Neurosciences \& Biomedical Engineering, Case Western Reserve University, Cleveland, OH, USA.
\textsuperscript{13}Inst.\ of Neuroinformatics, UZH-ETH Zurich, Zurich, Switzerland.
\textsuperscript{14}Dept.\ Control \& Dynamical Systems, Caltech, Pasadena, CA, USA.
\textsuperscript{15}Dept.\ Electrical \& Computer Engineering, UC Santa Cruz, Santa Cruz, CA, USA.
\textsuperscript{16}Dept.\ Brain and Cognitive Sciences, MIT, Cambridge, MA, USA.
\textsuperscript{19}Dept.\ Electrical \& Computer Engineering, University of Cincinnati, Cincinnati, OH, USA.
\textsuperscript{20}Dept.\ Engineering, Oakland University, Rochester, MI, USA.
\textsuperscript{21}Dept.\ Computer Science, Rochester Institute of Technology, Rochester, NY, USA.
\textsuperscript{22}Champalimaud Neuroscience Programme, Champalimaud Foundation, Lisbon, Portugal.
\textsuperscript{23}School of Computer Science $+$ Montreal Neurological Institute, McGill University \& Mila, The Quebec AI Institute.
\textsuperscript{24}Institute of Computational Life Sciences, Zurich University of Applied Sciences, W\"{a}denswil, Switzerland.
\textsuperscript{25}School of Electrical Engineering \& Computer Science, Penn State University, University Park, PA, USA.
\textsuperscript{26}Dept.\ Neurobiology, UC San Diego, La Jolla, CA, USA.
\textsuperscript{27}Dept.\ Physiology, UC San Francisco, San Francisco, CA, USA.
\textsuperscript{28}Cold Spring Harbor Laboratory, Cold Spring Harbor, NY, USA.
\textsuperscript{29}Dept.\ Computer Science and Biomedical Engineering, Yale University, New Haven, CT, USA.
\textsuperscript{30}Mathematics Department, Kalaheo High School, Kailua, HI, USA.
\noindent\textsuperscript{\dag}Corresponding authors:
Jean-Marc Fellous (\texttt{jmfellous@ucsd.edu});
Terrence Sejnowski (\texttt{terry@snl.salk.edu});
Anthony Zador (\texttt{zador@cshl.edu}).

\noindent\textit{Authors emails:}
Gina Adam (\texttt{ginaadam@email.gwu.edu});
James B Aimone (\texttt{jbaimon@sandia.gov});
Akwasi Akwaboah (\texttt{aakwabo1@jhu.edu});
Yiannis Aloimonos (\texttt{jyaloimo@umd.edu});
Carmen Amo Alonso (\texttt{camoalon@stanford.edu});
Chiara Bartolozzi (\texttt{chiara.bartolozzi@iit.it});
Michael J.\ Bennington (\texttt{mbenning@andrew.cmu.edu});
Michael Berry (\texttt{berry@princeton.edu});
Bing W.\ Brunton (\texttt{bbrunton@uw.edu});
Gert Cauwenberghs (\texttt{gcauwenberghs@ucsd.edu});
Hillel J.\ Chiel (\texttt{hjc@case.edu});
Tobi Delbruck (\texttt{tobi@ini.uzh.ch});
John Doyle (\texttt{doyle@caltech.edu});
Jason Eshraghian (\texttt{jsn@ucsc.edu});
Ralph Etienne-Cummings (\texttt{retienne@jhu.edu});
Cornelia Ferm\"{u}ller (\texttt{fermulcm@umd.edu});
Matthew Jacobsen (\texttt{mattjac@umd.edu});
Ali A.\ Minai (\texttt{ali.minai@uc.edu});
Barbara Oakley (\texttt{barbo8@gmail.com});
Alexander G.\ Ororbia II (\texttt{agovcs@rit.edu});
Joe Paton (\texttt{joe.paton@neuro.fchampalimaud.org});
Blake Richards (\texttt{blake.richards@mila.quebec});
Yulia Sandamirskaya (\texttt{sank@zhaw.ch});
Abhronil Sengupta (\texttt{sengupta@psu.edu});
Shihab Shamma (\texttt{sas@umd.edu});
Michael Stryker (\texttt{michael.stryker@ucsf.edu});
Seong Jong Yoo (\texttt{yoosj@umd.edu});
Steven W.\ Zucker (\texttt{steven.zucker@yale.edu}).

}
\vfill
\noindent\textit{Current draft date: \today}
\end{onecolumn}

\newpage
\begin{twocolumn}

\section{Introduction}

The major leaps in artificial intelligence (AI) share a common origin: neuroscience.
McCulloch and Pitts modeled the neuron as a logical gate; Rosenblatt
turned that model into a learning machine. Hubel and Wiesel mapped the
hierarchical organization of visual cortex \newcite{HubelWiesel1962}; Fukushima and later LeCun
translated it into convolutional networks \newcite{Fukushima1980,LeCunEtAl1989,LeCunEtAl1998}. The attention mechanisms at the core of modern
transformers draw on the cortical principle that processing resources
must be allocated selectively to relevant inputs \newcite{MoranDesimone1985,BahdanauChoBengio2015}. Each of these, and
many more, was a specific computational insight extracted from a
specific neuroscience finding, then engineered into an algorithm
\cite{hassabis2017neuroscience,ford2024exploring,zador2023catalyzing,marblestone2016toward} \newcite{summerfield2026, Lake_Ullman_Tenenbaum_Gershman_2017, Hwang_2026}.
The deep learning revolution descends from this lineage. \new{We use the term \textit{NeuroAI} for the bidirectional research program linking neuroscience and artificial intelligence: studying biological intelligence to extract computational principles that can advance AI, and using AI both as a tool for neuroscience and, where appropriate, as a model for understanding biological intelligence.}

Industry is now investing hundreds of billions of dollars per year in AI.
Frontier models improve rapidly. Large language models pass professional
examinations, write and debug functional code, \del{and} generate images and videos from natural
language descriptions, \new{and solve open problems in mathematics}. Some leaders in the field predict artificial
general intelligence within a few years. These achievements are
substantial, and scaling current approaches may well continue to produce
gains.

Nearly all of this investment, however, is concentrated on a single
architectural paradigm: transformers trained on massive static datasets,
running on dense synchronous digital GPU computation. The bet is also conceptually narrow:
\new{much of frontier AI is trained and evaluated primarily through language.
However, language represents only one particular class of intelligent behavior.
Nonhuman animals exhibit sophisticated learning, decision-making, navigation,
and control without language, and in humans, the language network is
anatomically dissociable from systems supporting other cognitive functions}
\cite{fedorenko2024language}\new{. The focus on language-centric
architectures therefore explores only a restricted part of the space of
possible mechanisms that underwrite intelligent behavior, and may miss capabilities that
benchmarks within the language domain do not reveal.} \new{These capabilities include affective modulation of behavior, unconscious inferences, flashes of insight or cognitive incubation.} \new{Many major advances
in AI---including perceptrons, convolutional networks, reinforcement learning,
and attention mechanisms---introduced qualitatively new computational ideas
rather than emerging from the scaling of existing approaches.} The question
facing the field now is whether the next
generation of architectural ideas will be ready when the current
paradigm reaches its limits.

Three fundamental capability gaps are already visible, and they are
unlikely to close through scaling alone. First, AI struggles to interact with
the physical world: it passes the bar exam but cannot clear a dinner
table and wash the dishes. Second, AI does not learn the way animals do---continuously,
adaptively, with graceful fallback under novelty---and this makes it brittle.
Third, AI is inefficient in energy and data to a degree that constrains
who can build it and where it can be deployed. These gaps are
architectural. They reflect deep mismatches between how current AI systems
are designed and how biological intelligence works. They are also the
problems that brains evolved to solve.

Neuroscience is now positioned to address them with a specificity that
was not previously possible. The complete connectome of the fruit fly brain has been mapped at synaptic resolution \newcite{dorkenwald2024flywire,scheffer2020hemibrain}. Large-scale
electrophysiology using Neuropixels probes allows simultaneous recording from thousands of neurons across multiple brain areas \newcite{Siegle2026LargeScale}. Optogenetic tools enable causal manipulation of identified cell
types \newcite{boyden2005optogenetics}. Comprehensive behavioral tracking links neural activity to behavior with a precision unavailable a generation ago. For the first time, we can extract specific algorithmic insights from biological computation about how neural circuits achieve robust \newcite{Theilman2026Intrinsic} , efficient, and adaptive performance.

We lay out three promising foundations for the next generation of AI.
We describe the three capability gaps in detail and identify the
neuroscience principles that address each one. We present a research
roadmap for the coming decade, organized by concrete milestones. And we
describe the institutional changes, in training, infrastructure, and
community organization, needed to realize this program. The work we
describe complements industry's current efforts, which are focused on optimizing
the present paradigm. The research program we propose here
develops the architectural alternatives that will be needed next,
whether the current paradigm plateaus in three years or in fifteen.
\new{Figure~\ref{fig:summary} summarizes the argument: it maps the
capability gaps of modern AI onto the neuroscience principles that
address them, and onto the outcomes that a neuroscience-informed program
could deliver.}

This paper emerged from a workshop convened by the National Science
Foundation in August 2025, bringing together about seventy
researchers and program managers from neuroscience, artificial intelligence, robotics,
neuromorphic engineering, and cognitive science
\cite{fellous2026neuroai}. The workshop was organized around five
thematic areas---embodied cognition and computation, language and
communication, robotics, learning in humans and machines, and
neuromorphic AI engineering---but the discussions repeatedly converged
on a smaller set of cross-cutting principles. We present those
convergences here \cite{Sejnowski2024Future}.

\begin{figure}[tb]
  \centering
  \includegraphics[width=\columnwidth]{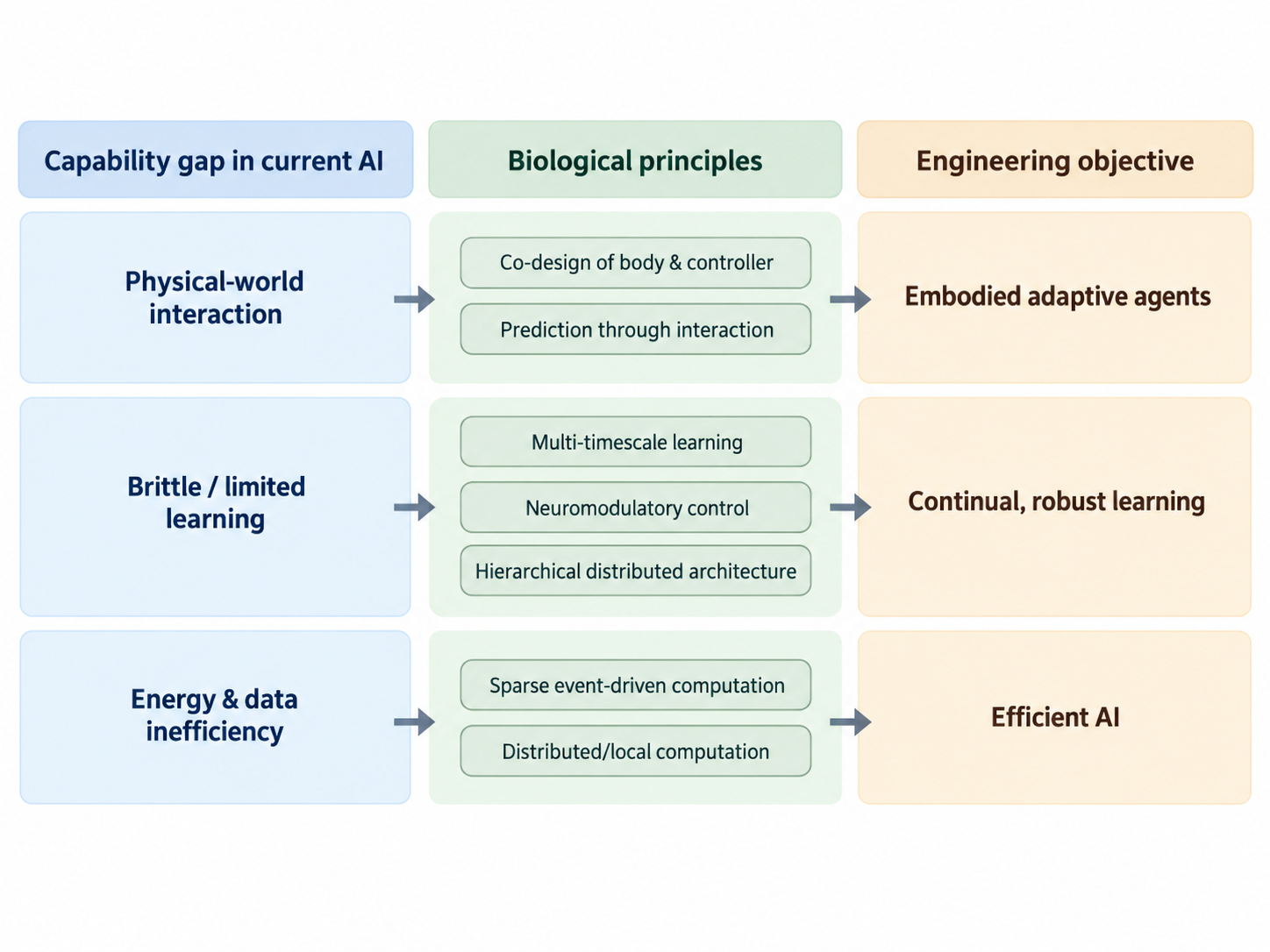}
  \caption{\textbf{Mapping capability gaps to biological principles and
  engineering objectives.} Each of the three capability gaps in current AI
  (left)---physical-world interaction, brittle and limited learning, and
  energy and data inefficiency---is addressed by a set of neuroscience
  principles (center): co-design of body and controller and prediction through
  interaction; multi-timescale learning, neuromodulatory control, and
  hierarchical distributed architecture; and sparse event-driven and
  distributed, local computation. Pursuing these principles points toward the
  corresponding engineering objectives (right): embodied adaptive agents,
  continual and robust learning, and efficient AI.}
  \label{fig:summary}
\end{figure}

\section{Three Problems}
\label{sec:three_problems}

We organize the limitations of current AI into three categories: the
limited ability to interact robustly with the physical world, a mode of learning that
is both inflexible and fragile, and an inefficiency in energy and data
that limits who can build AI and where it can operate. For each, we
describe the capability gap, explain why scaling current approaches is
unlikely to close it, and identify the neuroscience principles that
point toward a solution.

\subsection{AI struggles to interact with the physical world}
\label{sec:problem1_embodiment}

In 1988, Hans Moravec observed that the tasks humans find
difficult---chess, calculus, logical reasoning---are easy for computers,
while the tasks we find effortless---walking, catching a ball,
recognizing a face across a room---are extraordinarily hard
\cite{moravec1988mind}\new{~(Figure~\ref{fig:moravec})}. Nearly four decades later, the pattern holds in
a new form. AI systems can now pass medical licensing exams, generate
publishable prose, and write software. But decades into robotics
research, the most capable robot in a typical household is a vacuum
puck that bumps into furniture\new{, requires house preparation and regular rescuing}. No existing system can crack an egg,
separate the yolk from the white, and make an omelet in an unfamiliar
kitchen.

\begin{figure}[tb]
  \centering
  \includegraphics[width=\columnwidth]{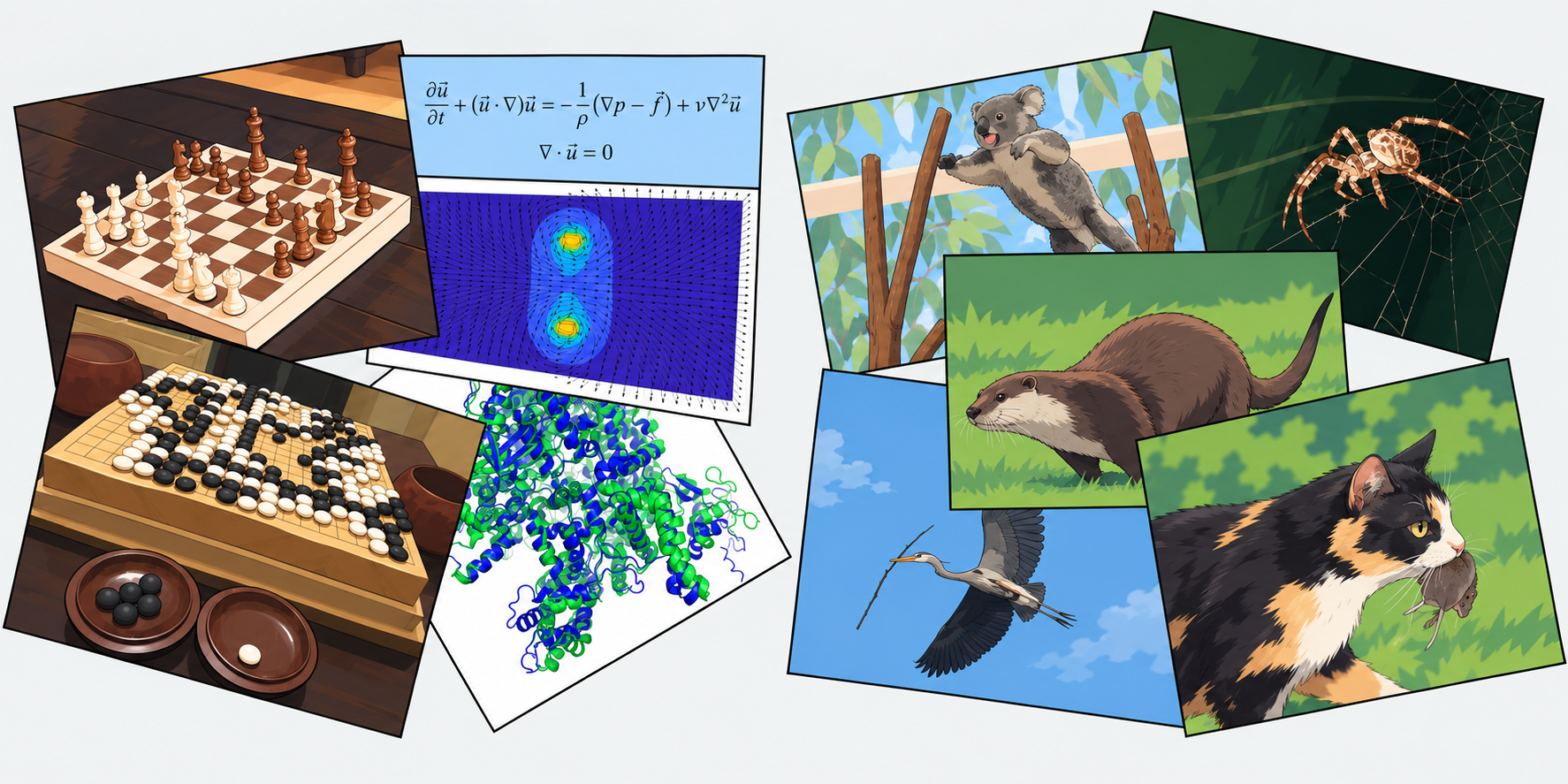}
  \caption{\textbf{Moravec's paradox, updated for modern AI.} AI has made rapid progress on tasks that humans regard as cognitively demanding—including games, mathematics, language, and scientific prediction—while remaining comparatively weak at the real-time sensing, control, and physical interaction that animals perform routinely. This contrast motivates the search for computational principles underlying embodied biological intelligence.}
  \label{fig:moravec}
\end{figure}

The reason is that physical interaction with the world poses a different
kind of computational problem from the ones current AI is optimized for.
Real-world action requires continuous control in real time, with noisy
and partial sensor data, physical consequences for errors, and no option
to reset. The agent must handle objects it has never seen, surfaces with
unknown friction, forces it did not expect. These are the conditions under which
biological intelligence evolved, and every animal on earth---from
insects navigating turbulent air to octopuses manipulating objects with
eight flexible arms---solves them routinely.

Current AI systems are trained on static datasets: text, images, video.
They observe the world passively, through data collected and curated by others. \new{Reinforcement-learning agents are a partial exception: systems such as deep Q-networks and AlphaZero build representations through their own interaction with a virtual environment rather than from a fixed corpus \newcite{mnih2015dqn,silver2018alphazero}. The dominant paradigm behind today's largest models nonetheless still learns from static data, and even interactive agents are typically confined to simulators or games rather than the open physical world.}
Whatever internal representations large language models (LLMs) develop
from this passive exposure, those representations were not built through
the agent's own causal interaction with physical reality. A child
learning the word ``heavy'' has spent months picking things up before
encountering the word. The meaning is anchored in proprioceptive and
tactile experience, and is augmented by auditory experience when the object falls. An LLM's representation of ``heavy'' is anchored in
the co-occurrence statistics of text.

Neuroscience offers two relevant constructive principles. The first is
\textit{co-design of body and controller}. Biological brain architectures
did not evolve in isolation. They co-evolved with the bodies they
control, producing tight coupling between morphology and neural
circuitry \cite{chiel1997brain}. A fly's visual system is tuned to the optic flow statistics
generated by its particular flight dynamics. An owl's auditory processing
is calibrated to the interaural time differences produced by its head
geometry. The computational problems the brain solves are shaped by the
body it inhabits \cite{dangelo2026benchmarking}.

This co-evolution has concrete engineering consequences.
Clever body design offloads computation from the neural
controller. Tendon-driven limbs exploit passive dynamics---springs and
dampers in the musculoskeletal system---to store and release energy
during locomotion, reducing the demands on the nervous system. A human
fingertip contains roughly 240 mechanoreceptors per square centimeter,
providing dense information about texture, pressure, slip, and
vibration \newcite{johansson1979tactile}. This massive peripheral sensing, integrated with
proprioception and vision through tightly coupled neural circuits, is
what makes human manipulation so adaptive. The retina illustrates the
same principle at a different scale: rather than transmitting raw
images, it performs edge detection, motion detection, contrast
normalization, and prediction of upcoming stimuli, compressing
\new{its output}\del{information} by orders of magnitude before it reaches the brain.
Neuromorphic sensors, particularly dynamic vision sensors that encode
\del{luminance} \new{brightness} changes asynchronously rather than capturing static frames,
represent an important engineered step toward this biological strategy.
The next step is to combine (as in biology) these transients
with the sustained \new{signals}\del{information} in frames to emulate the computations performed by
the retina and early visual cortex, enabling efficient algorithms and,
ultimately, hardware implementations that provide biologically grounded
primitives on which higher-level processes can build.

The physical structure of the body shapes the control problems it must
solve. The soft arms and suckers of the octopus are part of its solution
for handling complex materials; more generally, embodied intelligence
offloads part of the control problem onto the periphery. The size and
speed of a behavior alter the relevant dynamics, and biological nervous
systems adjust their control strategies accordingly
\cite{sutton2023phase}.

Modern robots violate the co-design principle almost entirely. Most are
rigid, over-actuated, and energy-inefficient, designed this way because
rigid bodies are simpler to model mathematically. The control system and
the physical hardware are typically designed by separate teams with
separate objectives. The result is machines capable of precise
repetitive motions in controlled environments but poor at adapting to
novel objects, surfaces, or forces.

The second principle is \textit{prediction through interaction}. The
brain is organized around prediction at every level of the sensory
hierarchy. Retinal ganglion cells respond most strongly when the visual
scene violates their expectations, transmitting prediction errors rather
than raw data \newcite{hosoya2005dynamic}. Predictive coding models \cite{salvatori2025survey,gunasekaran2025predictive}
propose that each level of the cortical hierarchy maintains a generative
model of its inputs, passing only the discrepancies upward.  \new{It has primary roots in the retina and visual cortex \newcite{srinivasan1982predictive,rao1999predictive, olshausen1996emergence}; it is an influential framework, but it remains debated, and the direct neurophysiological evidence for it is still contested \newcite{walsh2020evaluating}.} If this
framework is approximately correct, the brain's primary computational
strategy is to predict the next state of the world conditioned on its
own actions, and to learn by correcting those predictions. \new{ Such brain-like predictive coding can be efficient on neuromorphic hardware, and can be applied to solve real world problems \newcite{Theilman2025Sparse}.}

This has direct implications for how AI systems should be trained.
Infants learn in a developmental sequence: first building representations
through passive observation, then learning cause and effect by
incorporating their own actions as motor control matures, then
constructing causal models through active experimentation. This
progression---from passive prediction to action-conditioned prediction
to causal reasoning---suggests a natural curriculum for training
embodied AI agents, and aligns with current proposals like LeCun's
Joint Embedding Predictive Architecture \cite{assran2023self}, which
learns by predicting future states in abstract representation space
rather than in pixel or token space.

Physical robots are essential to this program. They are not just consumers
of NeuroAI principles; they generate the questions that sharpen them.
Building a robot forces precise definitions of otherwise vague concepts.
When a neural architecture designed for one body plan is applied to a
robot with a fundamentally different morphology, the resulting failures
expose assumptions about brain-body coupling that purely computational
work leaves untested. Robots reproduce the actual operating conditions of
biological intelligence---partial data, continuous control, sparse
reward, real consequences---in ways that curated benchmarks cannot.

Interacting with the physical world includes interacting with others,
be they humans or machines. \del{Current AI LLMs can emulate empathy, but do
not yet possess it, resulting in sometimes disastrous consequences.}\new{Current
LLMs can generate empathic-sounding language, but their subsequent actions do
not always reflect the concern or values expressed in those words, sometimes
with harmful consequences \newcite{Kumar2026}. Interestingly, recent work suggests that LLMs may be spontaneously evolving internal representations of some emotions \newcite{sofroniew2026emotionconceptsfunctionlarge} .}
Humans may be fooled into thinking the chatbot cares for them, when in
fact it does not. AI systems do not have a genuine sense of values, or
of right and wrong. They currently do not have artificial emotions; the
functional equivalent of human or animal emotions. Yet such emotions
serve many computational purposes including providing feedback that drives
learning \cite{arbib2004emotions}. A dog trained by a scolding tone modifies its behavior; a
presenter corrected in public feels embarrassment and avoids the same
mistake. Current LLMs apologize graciously when told they are wrong,
but cannot use the emotional consequences of their errors as a
learning signal to improve.

\subsection{AI learning is limited}
\label{sec:problem2_limited_learning}

\new{Most deployed }AI systems \del{cannot}\new{do not} learn from experience after \del{deployment}\new{training ends, despite active efforts in the continual-learning literature that seeks to close this gap \newcite{parisi2019continual,kirkpatrick2017overcoming}}. A physician
improves with every patient seen. A factory worker notices when a
machine starts sounding different and adjusts. A self-driving car that
encounters an unusual intersection \new{only learns from it when the fleet data is consolidated and redeployed} \del{does not learn from it}. A medical
imaging classifier trained at one hospital degrades when deployed at
another, where the scanners, patient demographics, and imaging protocols
differ, and it cannot self-correct. The model is generally frozen the
moment training ends. Whatever it knew at that point is all it will ever
know, until a successor model is retrained.

A related failure is that current systems do not represent what they do
not know. Hallucinations---the confident generation of false or
fabricated information---are the most visible symptom. When an LLM's
training data is sparse or contradictory on a topic, the system produces
fluent confabulation with no internal signal that anything is wrong. A
human who has never done plumbing knows they don't know plumbing.
Current AI has no comparable self-model of its own competence
boundaries, or of its confidence level in its answers, because it has no
episodic record of what it has and has not experienced and no grounded
world model against which to check its outputs.

Within the domain of language, frontier LLMs have become impressively
robust. They handle paraphrased questions, ambiguous prompts, and
unusual formatting with a fluency that would have seemed impossible five
years ago. The same is becoming true of image or audio generation.
Outside these domains, the picture is different. Self-driving
systems still fail at edge cases with fatal consequences. Robotic
manipulation breaks when objects differ modestly from training examples.
Industrial monitoring systems degrade when manufacturing conditions
drift. These are domains where the system encounters the world as it
actually is, rather than as a curated dataset represents it.

The field has recognized the \del{memory} problem \new{posed by the lack of an episodic memory} and responded with
engineering workarounds. \new{Elaborate human-engineered prompt injections and retrieval-augmented} \del{Retrieval-augmented} generation (RAG) gives
models access to external documents at inference time. Context windows
have grown from thousands to millions of tokens. Tool use lets models
query databases and APIs. Fine-tuning adapts models to new domains.
These help, but they are workarounds for an architectural gap, not
solutions to it. RAG retrieves documents but does not change what the
model knows in the long run. A longer context window is still a fixed
buffer, wiped between sessions. Fine-tuning can degrade previously learned
knowledge and is typically performed offline rather than continuously. None of these approaches give
the system the ability to update its internal model of the world in
response to ongoing experience. They simulate memory without
implementing learning.

These failures, the inability to learn after deployment, the absence of
calibrated uncertainty, the brittleness in physical domains, the silent
degradation, share a common architectural root: current AI has no
mechanism for continual, multi-timescale learning and no hierarchical
organization that separates fast safe responses from slow flexible ones.
There is no hierarchical organization separating detailed from `gist'
memory. Neuroscience offers specific principles for addressing both \cite{mathis2025leveraging} \newcite{summerfield2026}.

\subsubsection{Multiple memory systems operating at different timescales}
Biological learning spans a continuous spectrum: sensory habituation in
seconds, motor adaptation over minutes to hours, episodic and semantic
memory consolidation over days to years, evolutionary shaping of neural
architecture over millennia. Each timescale serves a distinct function,
and the interaction between them is what produces both adaptability and
stability. Long-term working memory used during cognitive processing lasts for hours but only salient and attended events are retained in long-term memory \cite{Sejnowski2026WorkingMemory}. The hippocampus learns rapidly, forming
sparse, pattern-separated representations of individual episodes within
one or a few exposures \newcite{yassa2011pattern}. The neocortex learns slowly, integrating across many episodes to extract statistical regularities. Sleep-based replay of
hippocampal activity transfers memories to cortical representations
gradually, enabling fast learning and stable long-term memory to coexist
because they are implemented in separate but communicating structures \newcite{wilson1994reactivation,diekelmann2010memory}.
Sleep allows for selective forgetting of likely useless information and
for the consolidation of likely important memory episodes and facts.
Both are important components of continuous lifelong learning. This is
the core insight of complementary learning systems theory \newcite{mcclelland1995complementary,kumaran2016complementary}, and
it provides an architectural solution to the stability-plasticity
dilemma that plagues all current approaches to continual learning in AI.

Neuromodulatory systems add context-sensitive control over this
architecture. Dopaminergic neurons broadcast reward prediction error
signals that gate plasticity in the striatum and prefrontal cortex \newcite{schultz1997neural}\new{, though recent work shows that these signals are more heterogeneous and region-specific than a single scalar prediction error \newcite{berke2018what,engelhard2019specialized}}.
Noradrenergic neurons in the locus coeruleus signal arousal and novelty,
shifting the balance between exploiting known strategies and exploring
alternatives \newcite{astonjones2005integrative}. Acetylcholine \del{levels control the timing and gating of synaptic plasticity, in addition to enabling muscle contraction and movement, crucial aspects of learning and memory in many tasks where behavioral actions are used to measure learning.}\new{modulates the timing and gating of synaptic plasticity in cortex and hippocampus, shaping attention, encoding, and memory consolidation \newcite{hasselmo2006role}.} These neuromodulatory signals
interact in ways that we are only beginning to
understand \cite{jang2026acetylcholine}. These signals specify whether
to learn and how much, operating differentially across brain regions and
synaptic populations, with a nuance that current optimization algorithms
do not approach \cite{schmidgall2024brain}\newcite{doya2002metalearning,mei2022informing}. \new{These neuroscience-inspired ideas have become concrete in modern reinforcement learning: dopamine has been reinterpreted through distributional codes for value \newcite{dabney2020distributional}, prefrontal cortex through meta-reinforcement learning \newcite{wang2018prefrontal}, and hippocampal codes through the successor representation \newcite{stachenfeld2017hippocampus}.} The spectrum of synaptic timescales provides a further
dimension: some synapses change in seconds through short-term
facilitation and depression, others require minutes of sustained
activity, still others change only after hours of protein-synthesis-dependent
consolidation. Fast synapses capture immediate statistical structure;
slow synapses accumulate evidence and resist transient fluctuations.
``Fast weights,'' transient synaptic modifications that store recent
context in associative memory, provide a neurally plausible mechanism
for algorithmic features such as the context window in a transformer \cite{Sejnowski2026WorkingMemory}\newcite{ba2016fastweights},
but one that emerges from synaptic dynamics and could be implemented in
neuromorphic hardware. Beyond weight changes, structural plasticity,
apoptosis, and adult neurogenesis modify the network's topology over
time, expanding representational capacity in a way that has no analogue
in current deep learning, where architecture is fixed at initialization.

\subsubsection{Hierarchical fallback architecture}
Robustness in biological systems arises from layered, distributed
control. The brain's motor system is a stack of specialized layers
operating at different timescales: the spinal cord mediates reflexes in
milliseconds, the cerebellum calibrates movements over seconds to
minutes, the basal ganglia select among competing action plans, and the
motor cortex and prefrontal cortex handle planning over longer horizons.
Each layer occupies a distinct neural structure with its own learning
rules and network dynamics.

The vestibulo-ocular reflex (VOR) illustrates how these layers interact.
The VOR stabilizes gaze during head movements with a latency of about
10 milliseconds, far too fast for cortical involvement. When the reflex
gain becomes incorrect (as when someone starts wearing new glasses), the
cerebellum detects the resulting retinal slip and slowly recalibrates
over hours to days; once the new gain is learned, it is transferred to
the vestibular nucleus for fast execution, freeing the cerebellum for
new calibration tasks \newcite{raymond1996cerebellum}. The general motif is slow adaptation followed by
fast deployment: invest time in learning, then execute what has been
learned with minimal latency.

John Doyle's framework of layered control architectures provides
mathematical grounding for why this pattern produces robustness
\cite{matni2024quantitative}. \new{The idea that robust behavior can emerge from layered controllers has deep roots in robotics, going back to Brooks's subsumption architecture \newcite{brooks1986subsumption}.} Architectures with hard constraints at
lower levels---what Doyle calls ``constraints that deconstrain''---are
the established path to reliable performance under uncertainty. Spinal
reflexes enforce safety constraints (withdrawal from pain, postural
correction) that free the cortex to plan without attending to
moment-by-moment stability. Distributed control extends this principle
throughout the motor system. Sensorimotor feedback loops operate at
every level of the hierarchy, with local controllers handling tasks like
joint compliance and reflex coordination that centralized robotic
systems must compute explicitly. Higher levels coordinate and set goals
rather than micromanaging execution. Recent advances in system-level
synthesis have made it possible to formally analyze such distributed
controllers with the sparse signals, time delays, and local feedback
characteristic of biological motor control, providing quantitative tools
for designing systems that are simultaneously distributed, adaptive, and
provably safe.

The engineering prescription is a hybrid: lower levels governed by
control theory with formal safety guarantees, higher levels governed by
learned policies that adapt to new tasks and environments. Current
robotic systems tend toward one extreme or the other, hand-designed
hierarchies that are reliable but inflexible, or end-to-end learned
controllers that are flexible but offer no safety guarantees. Biological
systems combine both. Getting this combination right is among the most
consequential open problems in NeuroAI.

\subsubsection{Starting from a better place}
One further principle addresses the inefficiency of current training and
supports all of the above. Current practice initializes deep networks
with random weights and trains to convergence: the equivalent of
building an organism from a \new{hand-crafted generic hyperarchitecture with no parameters} \del{random genome}. Biological development works
differently: the genome specifies a compact set of rules and a neural
architecture with multiple brain areas and an overall pattern of
connectivity, the result of uncountable ``training'' trials over
evolutionary history \cite{zador2019critique}. Within this framework the brain self-organizes
through developmental processes that interact with early sensory
experience. The resulting networks also contain useful priors, such as
edge detectors in visual cortex and tonotopic maps in auditory cortex,
before any task-specific learning begins. This is called inductive bias in machine learning. Replacing random
initialization with structured developmental programs that generate
useful starting points could bypass the need for exhaustive retraining
and immense data availability, and, more importantly, make continual adaptation more natural from the
outset \cite{minai2024deep}.

\subsection{AI is inefficient}
\label{sec:problem3_inefficiency}

Training a frontier language model consumes on the order of 50\,GWh of
electricity and costs hundreds of millions of dollars. A human brain
performs vastly more diverse computation on roughly 20\,W
\cite{chiel2026brain}. A child learns to recognize dogs from a handful
of examples; LLMs require effectively the entire internet. These
disparities determine where AI can be deployed, who can afford to build
it, and whether the technology's growth is sustainable. They also
concentrate AI capability in a small number of private organizations with access
to massive compute and data.

The root cause is architectural. Modern AI runs on von Neumann hardware,
in which memory and processing are physically separated. Moving data
between them consumes far more energy than the arithmetic itself. The
brain has no such bottleneck: synapses and neurons together store information and
perform computation, eliminating the cost of data transport. The
computational style also differs fundamentally. Transformers perform
dense matrix multiplications across billions of parameters at every step
regardless of input. Biological circuits are sparse: only \del{about 0.1\%}\new{a small fraction} of neurons are \new{spiking}\del{active} at any moment \newcite{lennie2003cost,attwell2001energybudget, zhou2025sparsity}, communication occurs
through discrete spikes, and the precise millisecond timing of those spikes carries
information that rate-based codes discard. Sparsity in the brain is a
computational advantage, yielding codes that are more energy-efficient,
more expressive per unit of activity, and more robust to noise. And
where current AI is predominantly feedforward, the brain is massively
recurrent: feedback connections from higher cortical areas to lower
sensory areas are at least as numerous as feedforward connections. This
feedback enables top-down prediction and iterative refinement; current
AI compensates for its absence by stacking more layers spatially rather
than reprocessing through fewer layers multiple times \cite{zhu2025looplm}, wasting both
memory and energy.

Neuromorphic computing implements these biological principles in
silicon. Chips like Intel's Loihi~2 run spiking neural networks in
which computation occurs only when a spike arrives; the hardware is
otherwise quiescent. The energy savings for sparse workloads are
substantial, but broader adoption requires algorithms designed natively
for event-driven processing rather than dense models retroactively
pruned \cite{eshraghian2023training, wunderlich2021event}. Bridging strategies are emerging: the ``spike as packet''
concept extends binary spikes to carry multibit payloads while
preserving sparse asynchronous communication, providing a path for
translating high-precision AI models onto neuromorphic substrates.
Rank-order coding and oscillation-based phase coding offer additional
schemes beyond simple rate codes.

Co-designing hardware and algorithms around sparsity is central to the
NeuroAI efficiency agenda \cite{zhou2025sparsity}. GPUs excel at dense synchronous computation
and are poorly matched to sparse, asynchronous, recurrent workloads.
Three-dimensional chip architectures that co-locate memory and
processing offer a further path forward: sparse computation generates
less heat per unit volume, potentially resolving the thermal problems
that prevent dense 3D integration. An immediate practical opportunity
also exists in commercial hardware. The neural processing units (NPUs) in
modern smartphones, optimized by fierce market competition for cost and
power efficiency, represent the state-of-the-art in battery-powered AI
processing. Opening these NPUs to applications beyond the smartphone
ecosystem, including robotics, prosthetics, and wearable health
monitoring, would give NeuroAI researchers access to powerful,
inexpensive platforms without waiting for custom neuromorphic chips to
reach production~scale~\cite{kudithipudi2025neuromorphic}.

\section{A Research Roadmap}
\label{sec:roadmap}

The principles described above are grounded in well-characterized
neuroscience, and several already have engineering implementations in
early stages. What has been missing is a coordinated program connecting
them. Here we outline research goals organized by the three problems
identified above, at three time horizons. We note that five-year
projections in fast-moving fields tend to be optimistic, while ten- to
twenty-year horizons tend to be conservative. The Human Genome Project
and the BRAIN Initiative both exceeded their long-term goals ahead of
schedule through sustained collaboration between scientists and
engineers.

\textbf{Connectome-based embodied digital twins} serve as a cross-cutting
platform for all three problems, providing a testbed for embodied
control strategies, learning architectures, and efficient neuromorphic
computation simultaneously. The fruit fly connectome, roughly 140,000
neurons and 50 million synapses, has been mapped \newcite{dorkenwald2024flywire} and initial simulations
have produced biologically plausible motor patterns. \new{Connectome-based agents have a long lineage, from early neuromechanical models of \textit{C.\ elegans} \newcite{izquierdo2013connectome,szigeti2014openworm} to recent connectome-constrained models of the fly \newcite{lappalainen2024connectome,shiu2024flybrain}. Experience with large-scale reconstruction and simulation efforts shows both the power and the limits of the approach. Anatomy is necessary but not sufficient: a mapped connectome does not by itself yield behavior, but when combined with physiology and modeling, it can strongly constrain circuit mechanisms and make previously intractable computations accessible, as work on the \textit{Drosophila} visual system has demonstrated \newcite{lappalainen2024connectome, Wang2026Memory}. For embodied digital twins, this circuit-level understanding must further be paired with an accurate body model, a biomechanical interface through which the nervous system acts on the world, and perhaps a Vision-Language-Action model (VLA) \newcite{Lake_Ullman_Tenenbaum_Gershman_2017, Kawaharazuka2025}.} Within five years,
a coarse-grained neuromorphic twin of the mouse brain (${\sim}$70 million
neurons) is feasible, providing a platform for testing how connectomic
structure supports learning and behavior. At the ten-year horizon, a
primate connectome twin would bring the field closer to a brain that
shares key architectural features with humans, including a six-layered
neocortex, a developed prefrontal cortex, and subcortical areas, with \del{abstracted circuit-level versions}\new{an abstracted circuit-level version} potentially usable for clinical applications in
disorders like Parkinson's disease and epilepsy. At the twenty-year
horizon, a functional digital twin of the human brain including the
peripheral nervous system would constitute both a landmark scientific
achievement and a source of architectural principles that cannot
currently be derived from first principles.

\textbf{Interacting with the physical world.} In the near term
(0--5 years), priorities include biomechanically realistic simulated
embodied agents in physics engines; robots with continual fine-tuning
for manipulation, terrain adaptation, and load management; event-based
neuromorphic sensing for real-time low-power robotic perception, with
foveated projection and event-based augmented reality as early
applications; and protective sensory channels---force, pressure,
temperature---that trigger withdrawal reflexes, giving robots the
equivalent of a pain response that prevents damage to themselves and
others. \new{Several of these directions are already active research programs rather than aspirations: learned legged locomotion now transfers to natural terrain \newcite{miki2022locomotion, Kumar2021-rma-legged-robots}, and reinforcement-learning controllers achieve champion-level drone racing \newcite{kaufmann2023drone}. The near-term goal is to consolidate and generalize such results under common principles and shared hardware.} In the mid-term (5--10 years), the emphasis shifts to world
models built from action-conditioned prediction \newcite{ha2018worldmodels,hafner2025dreamerv3,schrittwieser2020muzero}\new{, building on an already active world-model research program,} and validated on
physical robots rather than in simulation alone; fleet learning systems
in which many robots share experience and converge on solutions faster
than any individual\new{, a strategy already used at scale by autonomous-vehicle fleets}; and a standardized layered operating system for
robots, defining communication protocols between sensors, actuators,
local controllers, and higher-level planners. Foundational research on
the interaction of body and nervous system should proceed in parallel:
groups studying neuromechanics and motor control in animals should
collaborate closely with roboticists, so that robots serve as models
that sharpen biological questions and biological insights feed back into
peripheral design and neuroAI controllers. In the long term (10--20
years), the goal is to build robots with adaptive autonomy in
unstructured human environments---systems that handle novelty,
collaborate with people, and learn continuously from
interaction---supported by human-level tactile and olfactory sensing
with low-latency feedback and intrinsic ability to recognize and learn
the ethical and safety features of their actions. In animals and humans,
autonomy emerges from internal goals that shape perception and action.
Toddlers learn to avoid hot stoves and sharp corners, seek food and
comfort, and discover that inflicting pain on others brings
consequences. Integrating analogous internal goal states into artificial
controllers may be essential for exploration and learning in complex
environments without exhaustive training data.

\textbf{Robust learning.} Near-term goals include AI architectures with
multiple memory systems operating at different timescales and different
information resolution, testable as modules within existing frameworks;
incorporation of biological features such as neuromodulatory gating,
synaptic timescale diversity, and dendritic computation into hybrid
models; and community benchmarks for continual learning that test
graceful degradation under distributional shift, including but going
beyond resistance to catastrophic forgetting. In the mid-term,
priorities are hybrid hierarchical control architectures combining
formal safety guarantees at lower levels with learned flexibility at
higher levels; fully probabilistic AI systems that represent and
propagate uncertainty natively, essential for safety-critical domains
such as autonomous driving and surgical robotics; and developmental
initialization methods that replace random weights with structured
starting points generated through evolutionary or developmental search
over compact rule sets. Long-term goals include systems that
self-organize from genomic-like programs and already contain useful
priors before task-specific training begins; lifelong
agents and fleets that accumulate knowledge over years without degradation; and
closed-loop neuromorphic neural interfaces that sense, compute, and
stimulate in real time, the ultimate test of adaptive hierarchical
control.

\textbf{Efficiency.} Near-term priorities include community chip
development projects for hardware implementations of recurrent neural
networks and state-space models, architectures naturally suited to
neuromorphic substrates; repurposing smartphone NPUs as platforms for
edge NeuroAI applications in robotics, prosthetics, and wearable health
monitoring; neuromorphic wearable devices running spike-based inference
at ultra-low power; and software tools that evaluate algorithms against
hardware-aware efficiency metrics rather than accuracy alone. In the
mid-term, the agenda advances to heterogeneous 3D chip architectures
integrating distributed sensors, memory, and compute while exploiting
spatial, temporal, and precision-based sparsity; neuromorphic substrates
running models translated from high-precision AI via spike-as-packet and
related coding schemes; and bio-compatible neuromorphic interfaces
moving from laboratory demonstration to clinical prototype. The
long-term benchmark is a sub-kilowatt AI supercomputer: a system
performing at the level of today's large-scale clusters while consuming
less power than a household appliance. Achieving it will require every
principle described in this paper---sparse computation, \new{activity-driven} \del{event-driven}
processing, co-located memory and compute, 3D integration, and
algorithms designed for efficiency from the ground up---along with
neuromorphic chips manufactured at production scale.

\section{Enabling Conditions}

The roadmap above describes what should be built. Whether it gets built
depends on who does the work, what tools they have access to, and how
the research community is organized.

\textbf{Training the next generation.} NeuroAI requires researchers who
can move fluently between neuroscience and engineering, and there are not
enough of them. Most AI researchers are trained in computer science
departments with little exposure to neuroscience, cognitive science, or
control theory. Most neuroscientists cannot formalize the computational
principles they discover in ways that engineers can implement. The
result is a translation gap: important biological insights lie dormant
for years because no one in a position to implement them understands
their significance, and engineers draw on neuroscience at the level of
loose metaphor rather than specific mechanism.

Closing this gap requires structural changes to training at every career
stage, and the common principle is embedding: placing researchers in an unfamiliar
discipline long enough to acquire working competence, not just exposure.
Exposure to the brain, its computations and how it may inspire
\del{Artificial Intelligence}\new{artificial intelligence} can and should start in high school with
specifically designed AI-enabled devices and hands-on curricula suitable for this age-range. For
graduate students, this means interdisciplinary doctoral programs
designed around cross-training, including coursework in neuroscience,
cognitive science, dynamical systems, and control theory that goes
beyond a survey level, combined with research rotations in laboratories
outside the student's home discipline. For
postdoctoral researchers, it means fellowship programs that embed
AI-trained scientists in neuroscience laboratories for one to three
years, with a structured progression from applying AI tools to
neuroscience problems (which builds familiarity with the data and the
questions) toward carrying neuroscience principles back into AI
architecture and algorithm design. For established researchers, it means funded exchanges of weeks
to months in a collaborator's laboratory, with the explicit goal of
building a shared language and identifying joint problems. At every level, the key is
critical mass: isolated individuals placed in unfamiliar departments
rarely thrive, but cohorts of fellows at the same institution form a
community that sustains itself.

Programs like the Telluride, CapoCaccia, and Bangalore neuromorphic
workshops, which for decades have brought scientists
and technologists together on shared projects with support from NSF
and other agencies, provide
a proven model for shorter-duration immersion. Neuromatch and NeuroPAC
offer complementary approaches at different scales. \new{These programs have a substantial track record: they have collectively trained a large share of the current neuromorphic and computational-neuroscience workforce, and recent work analyzing NeuroAI training draws on this experience to recommend how to scale such pipelines across career stages \newcite{luppi2024trainees}.} Expanding and
sustaining programs like these across the full range of career stages is
among the most cost-effective investments the field can make. And because
the most important problems on this roadmap require teams, including
experimentalists, theorists, engineers, and roboticists,
funding mechanisms should support multi-laboratory collaborations rather
than individual investigators working in isolation.

\textbf{Hardware and infrastructure access.} Academic researchers in
neuromorphic computing face a specific bottleneck: the memory interface
and fabrication infrastructure needed to test their ideas at realistic
scale is proprietary, expensive, or unavailable. Open-source DRAM
interface IP, developed through joint effort between funding agencies
and industry, would enable academic groups to build neuromorphic chips
at modest cost relative to its impact. A successor to the MOSIS program---which from the 1980s
through the 2010s gave university researchers affordable access to chip fabrication and
enabled an entire generation of neuromorphic engineers---would have
comparable leverage if updated for
modern process nodes and 3D integration. More broadly, as industry
retires GPU clusters in favor of newer hardware, making that compute
available to academic researchers through national partnerships would
provide substantial capacity at low marginal cost for NeuroAI workloads
that are smaller than frontier LLM training but larger than typical
academic jobs.

\textbf{Standards and shared platforms.} Progress in computing has
always depended on common infrastructure. The x86 and ARM instruction
sets, the Linux and ROS operating systems, and the USB protocol each enabled
diverse innovation by providing stable layers that researchers could
build upon. NeuroAI lacks equivalents. A standardized layered
architecture for robots, defining communication protocols between
sensors, actuators, local controllers, and planning modules, would let
researchers swap components without redesigning entire systems. Shared
data repositories and standardized model formats, following the examples
of FASTA in genomics and Neurodata Without Borders in neuroscience,
would reduce duplication of effort that currently results from
incompatible simulation and modeling frameworks.

\textbf{Ethics and safety.} NeuroAI systems that interact with the
physical world, including robotic assistants, prosthetics, and
brain-computer interfaces, introduce safety and liability considerations that go beyond those posed by purely digital AI. Systems designed to model emotional processing or
social cognition raise questions about the potential for manipulation of
human users. Systems that adapt their behavior through continual learning pose novel
challenges for certification and regulation. Those that can physically act in the real world pose complementary challenges. These questions require
attention now, while the community is small enough and the technology
early enough for norms and safeguards to be established by the researchers building the
systems. Integrating AI ethics into training programs and engaging
proactively with policymakers are necessary and should not be deferred.

\textbf{The competitive landscape.} Industry AI laboratories command
resources, including compute, data, and engineering talent, that
academic groups cannot match. Industry's incentives, understandably,
favor optimizing the current paradigm over exploring alternatives. The international dimension adds urgency: substantial
government investments in AI, robotics, and neuroscience are being made
worldwide, with implications for scientific leadership and national
competitiveness.

Every major architectural transition in AI originated in
research environments where scientists had the freedom to explore
alternatives to the dominant paradigm of their day. \new{That freedom has, at
times, flourished in industry---notably during the golden age of Bell Labs and
IBM Research, and more recently at DeepMind---but industry is often under
greater pressure to exploit and scale successful approaches rather than explore
uncertain alternatives.} If the next architectural transition
follows this pattern, and there is no reason to think it will not, then
sustaining the academic research ecosystem that produces foundational
ideas is a strategic investment. NeuroAI is where the next set of ideas
is most likely to originate. Supporting it is supporting the future of
the field.

\section{Conclusion}

Artificial intelligence has explored a narrow region of the space of
possible architectures, one built on dense, feedforward,
energy-intensive computation that separates training from deployment and
ignores the physical world. Neuroscience has only recently acquired the
tools to characterize biological computation at the resolution needed to
extract engineering principles rather than loose analogies. The
convergence of these two trajectories defines the present opportunity.

We have identified three capability gaps in current AI---limited ability to
interact robustly with the physical world, inadequate learning that produces
brittle systems, and unsustainable inefficiency---and described the
neuroscience principles that address each one: co-design of body and
controller, prediction through interaction, multi-scale learning with
neuromodulatory control, hierarchical distributed architectures, and
sparse event-driven computation. Each corresponds to biological
mechanisms whose computational logic is increasingly understood, and
each points toward implementations that are feasible within the coming
decade. The research roadmap we have presented
translates these principles into concrete milestones, from neuromorphic
digital twins and community hardware platforms in the near term to
sub-kilowatt AI supercomputers and developmental self-organizing
architectures in the longer term.

Realizing this program requires training researchers who can work across
the boundary between neuroscience and engineering, hardware
infrastructure accessible to academic groups, community standards for
neuromorphic and robotic platforms, and sustained investment at a time
when the dominance of a single AI paradigm and the concentration of
resources in industry create strong headwinds for alternative
approaches.

The industrial revolution enhanced our physical capabilities. A
NeuroAI-informed transformation of artificial intelligence could
substantially enhance our cognitive capabilities while deepening our
understanding of our own minds. The researchers entering this field now will inherit experimental tools
and computational resources that previous generations could not have
imagined. The highest-leverage commitment that funders, universities,
and governments can make is to invest in their training, in the infrastructure
they need, and in the fundamental research that connects brains to
machines. NeuroAI is the continuation of the intellectual tradition that
produced modern AI. The next chapter depends on whether we choose to
sustain it.

\section*{Acknowledgements}

This manuscript is adapted from an NSF Workshop report entitled
``NeuroAI and Beyond'' held on August 27--29, 2025 in Arlington,
Virginia (Organizers: Cauwenberghs, Fellous, Ferm\"{u}ller,
Sandamirskaya, and Sejnowski). We are grateful to H.\,Chiel,
B.\,Oakley, C.\,Bartolozzi, B.\,Richards, E.\,Fedorenko, and T.\,Delbruck for leading
working-group discussions. We thank Drs.\ S.-S.\,Lim, E.\,Chua (NSF/SBE), and
G.\,Hwang (NINDS/BRAIN) for helpful discussions. \new{Images for Figure2 modified using GIMP under Creative Commons licenses. Original image credit: Chess board, Conal Gallagher; Go board, Zizou Man; Otter, Charlie from Bristol; Spider, photochemPA; Cat, Stig Nygaard; Heron, PhotoBobil (all from Wikimedia, (CC-BY 2.0)); Koala, Rainbow606 (Wikimedia, CC-BY 3.0). Protein folding \cite{jumper2021}, reproduced under a CC-BY 4.0.}

\section*{Author contributions}

J.-M.F., A.Z., and T.S. drafted and prepared the manuscript with assistance from Claude (Anthropic).  All authors edited, reviewed, and approved the final text.

\section*{Competing interests}

The authors declare no competing interests.

\bibliographystyle{unsrt}
\bibliography{neuroai,new-references-revision-2}

@article{arbib2004emotions,
  title={Emotions: from brain to robot},
  author={Arbib, Michael A and Fellous, Jean-Marc},
  journal={Trends in cognitive sciences},
  volume={8},
  number={12},
  pages={554--561},
  year={2004},
  publisher={Elsevier}
}

@article{wunderlich2021event,
  title={Event-based backpropagation can compute exact gradients for spiking neural networks},
  author={Wunderlich, Timo C and Pehle, Christian},
  journal={Scientific Reports},
  volume={11},
  number={1},
  pages={12829},
  year={2021},
  publisher={Nature Publishing Group UK London}
}

@article{chiel1997brain,
  author  = {Hillel J. Chiel and Randall D. Beer},
  title   = {The brain has a body: adaptive behavior emerges from interactions of nervous system, body and environment},
  journal = {Trends in Neurosciences},
  volume  = {20},
  number  = {12},
  pages   = {553--557},
  year    = {1997},
  doi     = {10.1016/S0166-2236(97)01149-1}
}

@article{minai2024deep,
  author  = {Ali A. Minai},
  title   = {Deep intelligence: what {AI} should learn from nature's imagination},
  journal = {Cognitive Computation},
  volume  = {16},
  pages   = {2389--2404},
  year    = {2024},
  doi     = {10.1007/s12559-023-10124-9}
}

@article{mathis2025leveraging,
  author  = {MacKenzie W. Mathis},
  title   = {Leveraging insights from neuroscience to build adaptive artificial intelligence},
  journal = {Nature Neuroscience},
  volume  = {29},
  pages   = {13--24},
  year    = {2026},
  doi     = {10.1038/s41593-025-02169-w}
}

@article{zador2019critique,
  author  = {Anthony M. Zador},
  title   = {A critique of pure learning and what artificial neural networks can learn from animal brains},
  journal = {Nature Communications},
  volume  = {10},
  number  = {1},
  pages   = {3770},
  year    = {2019},
  doi     = {10.1038/s41467-019-11786-6}
}

@article{zhou2025sparsity,
  author  = {Sheng Zhou and Chang Gao and Tobi Delbruck and Marian Verhelst and Shih-Chii Liu},
  title   = {Exploiting neuro-inspired dynamic sparsity for energy-efficient intelligent perception},
  journal = {Nature Communications},
  volume  = {16},
  pages   = {9928},
  year    = {2025},
  doi     = {10.1038/s41467-025-65387-7}
}

@article{dangelo2026benchmarking,
  author  = {Gianfranco D'Angelo and Jens E. Pedersen and Tarek Hassan and others},
  title   = {A benchmarking framework for embodied neuromorphic agents},
  journal = {Nature Machine Intelligence},
  volume  = {8},
  pages   = {300--312},
  year    = {2026},
  doi     = {10.1038/s42256-026-01197-w}
}

@article{eshraghian2023training,
  author  = {Jason K. Eshraghian and Max Ward and Emre O. Neftci and Xinxin Wang and Gregor Lenz and Girish Dwivedi and Mohammed Bennamoun and Doo Seok Jeong and Wei D. Lu},
  title   = {Training spiking neural networks using lessons from deep learning},
  journal = {Proceedings of the IEEE},
  volume  = {111},
  number  = {9},
  pages   = {1016--1054},
  year    = {2023},
  doi     = {10.1109/JPROC.2023.3308088}
}

@article{schmidgall2024brain,
  author  = {Samuel Schmidgall and Rojin Ziaei and Jascha Achterberg and Thomas Miconi and Louis Kirsch and S. Pardis Hajiseyedrazi and Jason Eshraghian},
  title   = {Brain-inspired learning in artificial neural networks: A review},
  journal = {APL Machine Learning},
  volume  = {2},
  number  = {2},
  pages   = {021501},
  year    = {2024},
  doi     = {10.1063/5.0186054}
}

@article{gunasekaran2025predictive,
  author  = {Skye Gunasekaran and Assel Kembay and Hugo Ladret and Rui-Jie Zhu and Laurent Perrinet and Omid Kavehei and Jason Eshraghian},
  title   = {A predictive approach to enhance time-series forecasting},
  journal = {Nature Communications},
  volume  = {16},
  pages   = {8645},
  year    = {2025},
  doi     = {10.1038/s41467-025-63786-4}
}

@article{zhu2025looplm,
  author  = {Rui-Jie Zhu and Zixuan Wang and Kai Hua and others},
  title   = {Scaling latent reasoning via looped language models},
  journal = {arXiv preprint arXiv:2510.25741},
  year    = {2025}
}

@article{sutton2023phase,
 author = {Sutton, Gregory P and Szczecinski, Nicholas S and Quinn, Roger D and Chiel, Hillel J},
 journal = {PNAS nexus},
 number = {10},
 pages = {pgad298},
 publisher = {Oxford University Press US},
 title = {Phase shift between joint rotation and actuation reflects dominant forces and predicts muscle activation patterns},
 volume = {2},
 year = {2023}
}

@article{hassabis2017neuroscience,
 author = {Hassabis, Demis and Kumaran, Dharshan and Summerfield, Christopher and Botvinick, Matthew},
 journal = {Neuron},
 number = {2},
 pages = {245--258},
 publisher = {Elsevier},
 title = {Neuroscience-inspired artificial intelligence},
 volume = {95},
 year = {2017}
}

@book{ford2024exploring,
 author = {Ford, Celia and Childers, Eva and Norris, Sheena M Posey},
 title = {Exploring the Bidirectional Relationship Between Artificial Intelligence and Neuroscience: Proceedings of a Workshop},
 publisher = {The National Academies Press},
 address = {Washington, DC},
 year = {2024},
 doi = {10.17226/27764}
}

@article{zador2023catalyzing,
 author = {Zador, Anthony and Escola, Sean and Richards, Blake and {\"O}lveczky, Bence and Bengio, Yoshua and Boahen, Kwabena and Botvinick, Matthew and Chklovskii, Dmitri and Churchland, Anne and Clopath, Claudia and others},
 journal = {Nature communications},
 number = {1},
 pages = {1597},
 publisher = {Nature Publishing Group UK London},
 title = {Catalyzing next-generation artificial intelligence through neuroai},
 volume = {14},
 year = {2023}
}

@article{fedorenko2024language,
 author = {Fedorenko, Evelina and Ivanova, Anna A and Regev, Tamar I},
 journal = {Nature Reviews Neuroscience},
 number = {5},
 pages = {289--312},
 publisher = {Nature Publishing Group UK London},
 title = {The language network as a natural kind within the broader landscape of the human brain},
 volume = {25},
 year = {2024}
}

@misc{fellous2026neuroai,
 author = {Jean-Marc Fellous and Gert Cauwenberghs and Cornelia Ferm{\"u}ller and Yulia Sandamirskaya and Terrence J. Sejnowski},
 title = {{NeuroAI} and Beyond},
 howpublished = {Preprint at \url{https://arxiv.org/abs/2601.19955}},
 year = {2026}
}

@book{moravec1988mind,
 author = {Moravec, Hans P.},
 title = {Mind Children: The Future of Robot and Human Intelligence},
 publisher = {Harvard University Press},
 year = {1988}
}

@article{salvatori2025survey,
 author = {Salvatori, Tommaso and Mali, A. and Buckley, Christopher L. and Lukasiewicz, Thomas and Rao, Rajesh P. N. and Friston, Karl J. and Ororbia, Alexander},
 journal = {Neural networks},
 pages = {108161},
 title = {A survey on neuro-mimetic deep learning via predictive coding},
 volume = {195},
 year = {2025}
}

@inproceedings{assran2023self,
 author = {Assran, Mahmoud and Duval, Quentin and Misra, Ishan and Bojanowski, Piotr and Vincent, Pascal and Rabbat, Michael and LeCun, Yann and Ballas, Nicolas},
 booktitle = {Proceedings of the IEEE/CVF conference on computer vision and pattern recognition},
 pages = {15619--15629},
 title = {Self-supervised learning from images with a joint-embedding predictive architecture},
 year = {2023}
}

@article{jang2026acetylcholine,
 author = {Jang, Hee Jae and McMahon Ward, Royall and Golden, Carla EM and Constantinople, Christine M},
 journal = {Nature Neuroscience},
 pages = {1--11},
 publisher = {Nature Publishing Group US New York},
 title = {Acetylcholine demixes heterogeneous dopamine signals for learning and moving},
 year = {2026}
}

@article{matni2024quantitative,
 author = {Matni, Nikolai and Ames, Aaron D. and Doyle, John C.},
 journal = {IEEE Control Systems},
 pages = {52--94},
 title = {A Quantitative Framework for Layered Multirate Control: Toward a Theory of Control Architecture},
 volume = {44},
 year = {2024}
}

@article{chiel2026brain,
 author = {Chiel, Hillel J and Coggan, Jay S and Datta, Gourav and Fellous, Jean-Marc and Nourse, William RP and Quinn, Roger D and Thomas, Peter J},
 journal = {Biological cybernetics},
 number = {2},
 pages = {5},
 publisher = {Springer},
 title = {Brain-inspired energy efficient technologies for next-generation artificial intelligence},
 volume = {120},
 year = {2026}
}

@misc{Sejnowski2026WorkingMemory,
 author = {Terrence J. Sejnowski},
 title = {Dynamical Mechanisms for Coordinating Long-term Working Memory Based on the Precision of Spike-timing in Cortical Neurons},
 howpublished = {Preprint at
 \url{https://arXiv:2512.15891}},
 year = {2025}
}

@book{Sejnowski2024Future,
 author = {Terrence J. Sejnowski},
 title = {ChatGPT and the Future of AI},
 publisher = {MIT Press},
 year = {2024}
}

@article{marblestone2016toward,
  author  = {Adam H. Marblestone and Greg Wayne and Konrad P. Kording},
  title   = {Toward an integration of deep learning and neuroscience},
  journal = {Frontiers in Computational Neuroscience},
  volume  = {10},
  pages   = {94},
  year    = {2016},
  doi     = {10.3389/fncom.2016.00094}
}

@article{kudithipudi2025neuromorphic,
  author  = {Dhireesha Kudithipudi and Catherine D. Schuman and Craig M. Vineyard and others},
  title   = {Neuromorphic computing at scale},
  journal = {Nature},
  volume  = {637},
  pages   = {801--812},
  year    = {2025},
  doi     = {10.1038/s41586-024-08253-8}
}

@Article{summerfield2026,
author={Summerfield, Christopher
and Stachenfeld, Kimberly},
title={The recipe for intelligence in natural and artificial systems},
journal={Neuron},
year={2026},
publisher={Elsevier},
issn={0896-6273},
doi={10.1016/j.neuron.2026.08.013},
url={https://doi.org/10.1016/j.neuron.2026.08.013}
}

@article{Lake_Ullman_Tenenbaum_Gershman_2017, 
title={Building machines that learn and think like people}, volume={40}, DOI={10.1017/S0140525X16001837}, journal={Behavioral and Brain Sciences}, author={Lake, Brenden M. and Ullman, Tomer D. and Tenenbaum, Joshua B. and Gershman, Samuel J.}, year={2017}, pages={e253}}

@INPROCEEDINGS{Kumar2021-rma-legged-robots, 
 title = "{RMA}: Rapid Motor Adaptation for Legged Robots", 
 author = "Kumar, Ashish and Fu, Zipeng and Pathak, Deepak and Malik, 
 Jitendra", 
 booktitle = "Robotics: Science and System XVII (RSS 2021)", 
 publisher = "RSS Foundation", 
 month = jul, 
 year = 2021, 
 url = "https://ashish-kmr.github.io/rma-legged-robots/", 
 doi = "10.15607/RSS.2021.XVII.011" 
}

@ARTICLE{Kawaharazuka2025,
  author={Kawaharazuka, Kento and Oh, Jihoon and Yamada, Jun and Posner, Ingmar and Zhu, Yuke},
  journal={IEEE Access}, 
  title={Vision-Language-Action Models for Robotics: A Review Towards Real-World Applications}, 
  year={2025},
  volume={13},
  number={},
  pages={162467-162504},
  doi={10.1109/ACCESS.2025.3609980}}

@Article{Kumar2026,
author={Kumar, Aakriti
and Poungpeth, Nalin
and Yang, Diyi
and Farrell, Erina
and Lambert, Bruce L.
and Groh, Matthew},
title={When large language models are reliable for judging empathic communication},
journal={Nature Machine Intelligence},
year={2026},
month={Feb},
day={01},
volume={8},
number={2},
pages={173-185},
issn={2522-5839},
doi={10.1038/s42256-025-01169-6},
url={https://doi.org/10.1038/s42256-025-01169-6}
}

@inproceedings{Wang2026Memory,
  author    = {Wang, Felix and Theilman, Bradley and Rothganger, Fred and Severa, William and  Vineyar, Craig and Aimone, James },
  title     = {Memory Trade-Offs in Neuromorphic Communication Strategies of the FlyWire Connectome on Loihi 2},
  booktitle = {2026 Neuro Inspired Computational Elements (NICE)},
  publisher = {IEEE},
  year      = {2026}
}

@inproceedings{Theilman2026Intrinsic,
  author    = {Theilman, Bradley H. and Aimone, James B.},
  title     = {Intrinsic Numerical Robustness and Fault Tolerance in a Neuromorphic Algorithm for Scientific Computing},
  booktitle = {2026 Neuro Inspired Computational Elements (NICE)},
  publisher = {IEEE},
  year      = {2026}
}

@article{Theilman2025Sparse,
  author  = {Theilman, Bradley H. and Aimone, James B.},
  title   = {Solving sparse finite element problems on neuromorphic hardware},
  journal = {Nature Machine Intelligence},
  volume  = {7},
  number  = {11},
  year    = {2025},
  pages   = {1845--1857}
}

@misc{sofroniew2026emotionconceptsfunctionlarge,
  
  author        = {Sofroniew, Nicholas and
                   Kauvar, Isaac and
                   Saunders, William and
                   Chen, Runjin and
                   Henighan, Tom and
                   Hydrie, Sasha and
                   Citro, Craig and
                   Pearce, Adam and
                   Tarng, Julius and
                   Gurnee, Wes and
                   Batson, Joshua and
                   Zimmerman, Sam and
                   Rivoire, Kelley and
                   Fish, Kyle and
                   Olah, Chris and
                   Lindsey, Jack},
title         = {Emotion Concepts and their Function in a Large Language Model},
howpublished = {Preprint at \url{https://arxiv.org/abs/2604.07729}},
year          = {2026}
  
}

@article{olshausen1996emergence,
  title={Emergence of simple-cell receptive field properties by learning a sparse code for natural images},
  author={Olshausen, Bruno A and Field, David J},
  journal={Nature},
  volume={381},
  number={6583},
  pages={607--609},
  year={1996},
  publisher={Nature Publishing Group UK London}
}

@article{Hwang_2026,
doi = {10.1088/1741-2552/ae8d5b},
url = {https://doi.org/10.1088/1741-2552/ae8d5b},
year = {2026},
month = {aug},
publisher = {IOP Publishing},
volume = {23},
number = {4},
pages = {043001},
author = {Hwang, Grace M and Falcone, Jessica D and Monaco, Joseph D and Pinard, Courtney R and Mollick, Jessica A and Miller, Roger L and Gage, Stephanie L and Kanaev, Andrey V and Kim, Margaret and Lukaszew, R Ale and Zehnder, Steven M and Rampulla, David},
title = {The role of neuromorphic principles in the future of biomedicine and healthcare},
journal = {Journal of Neural Engineering}
}

@article{Jumper2021, 
author = {John Jumper and Richard Evans and Alexander Pritzel and Tim Green and Michael Figurnov and Olaf Ronneberger and Kathryn Tunyasuvunakool and Russ Bates and Augustin Žídek and Anna Potapenko and Alex Bridgland and Clemens Meyer and Simon A.A. Kohl and Andrew J. Ballard and Andrew Cowie and Bernardino Romera-Paredes and Stanislav Nikolov and Rishub Jain and Jonas Adler and Trevor Back and Stig Petersen and David Reiman and Ellen Clancy and Michal Zielinski and Martin Steinegger and Michalina Pacholska and Tamas Berghammer and Sebastian Bodenstein and David Silver and Oriol Vinyals and Andrew W. Senior and Koray Kavukcuoglu and Pushmeet Kohli and Demis Hassabis}, doi = {10.1038/s41586-021-03819-2}, issn = {14764687}, issue = {7873}, journal = {Nature}, month = {8}, pages = {583-589}, pmid = {34265844}, publisher = {Nature Research}, title = {Highly accurate protein structure prediction with AlphaFold}, volume = {596}, year = {2021} }

@article{schultz1997neural,
  author  = {Schultz, Wolfram and Dayan, Peter and Montague, P. Read},
  title   = {A Neural Substrate of Prediction and Reward},
  journal = {Science},
  year    = {1997},
  volume  = {275},
  number  = {5306},
  pages   = {1593--1599},
  doi     = {10.1126/science.275.5306.1593}
}

@article{berke2018what,
  author  = {Berke, Joshua D.},
  title   = {What Does Dopamine Mean?},
  journal = {Nature Neuroscience},
  year    = {2018},
  volume  = {21},
  number  = {6},
  pages   = {787--793},
  doi     = {10.1038/s41593-018-0152-y}
}

@article{engelhard2019specialized,
  author  = {Engelhard, Ben and Finkelstein, Joel and Cox, Julia and Fleming, Weston and Jang, Hee Jae and Ornelas, Sharon and Koay, Sue Ann and Thiberge, Stephan Y. and Daw, Nathaniel D. and Tank, David W. and Witten, Ilana B.},
  title   = {Specialized Coding of Sensory, Motor and Cognitive Variables in {VTA} Dopamine Neurons},
  journal = {Nature},
  year    = {2019},
  volume  = {570},
  number  = {7762},
  pages   = {509--513},
  doi     = {10.1038/s41586-019-1261-9}
}

@article{astonjones2005integrative,
  author  = {Aston-Jones, Gary and Cohen, Jonathan D.},
  title   = {An Integrative Theory of Locus Coeruleus-Norepinephrine Function: Adaptive Gain and Optimal Performance},
  journal = {Annual Review of Neuroscience},
  year    = {2005},
  volume  = {28},
  pages   = {403--450},
  doi     = {10.1146/annurev.neuro.28.061604.135709}
}

@article{hasselmo2006role,
  author  = {Hasselmo, Michael E.},
  title   = {The Role of Acetylcholine in Learning and Memory},
  journal = {Current Opinion in Neurobiology},
  year    = {2006},
  volume  = {16},
  number  = {6},
  pages   = {710--715},
  doi     = {10.1016/j.conb.2006.09.002}
}

@article{mei2022informing,
  author  = {Mei, Jie and Muller, Eilif and Ramaswamy, Srikanth},
  title   = {Informing Deep Neural Networks by Multiscale Principles of Neuromodulatory Systems},
  journal = {Trends in Neurosciences},
  year    = {2022},
  volume  = {45},
  number  = {3},
  pages   = {237--250},
  doi     = {10.1016/j.tins.2021.12.008}
}

@article{doya2002metalearning,
  author  = {Doya, Kenji},
  title   = {Metalearning and Neuromodulation},
  journal = {Neural Networks},
  year    = {2002},
  volume  = {15},
  number  = {4--6},
  pages   = {495--506},
  doi     = {10.1016/S0893-6080(02)00044-8}
}

@article{dabney2020distributional,
  author  = {Dabney, Will and Kurth-Nelson, Zeb and Uchida, Naoshige and Starkweather, Clara Kwon and Hassabis, Demis and Munos, R{\'e}mi and Botvinick, Matthew},
  title   = {A Distributional Code for Value in Dopamine-Based Reinforcement Learning},
  journal = {Nature},
  year    = {2020},
  volume  = {577},
  number  = {7792},
  pages   = {671--675},
  doi     = {10.1038/s41586-019-1924-6}
}

@article{wang2018prefrontal,
  author  = {Wang, Jane X. and Kurth-Nelson, Zeb and Kumaran, Dharshan and Tirumala, Dhruva and Soyer, Hubert and Leibo, Joel Z. and Hassabis, Demis and Botvinick, Matthew},
  title   = {Prefrontal Cortex as a Meta-Reinforcement Learning System},
  journal = {Nature Neuroscience},
  year    = {2018},
  volume  = {21},
  number  = {6},
  pages   = {860--868},
  doi     = {10.1038/s41593-018-0147-8}
}

@article{stachenfeld2017hippocampus,
  author  = {Stachenfeld, Kimberly L. and Botvinick, Matthew M. and Gershman, Samuel J.},
  title   = {The Hippocampus as a Predictive Map},
  journal = {Nature Neuroscience},
  year    = {2017},
  volume  = {20},
  number  = {11},
  pages   = {1643--1653},
  doi     = {10.1038/nn.4650}
}

@article{mcclelland1995complementary,
  author  = {McClelland, James L. and McNaughton, Bruce L. and O'Reilly, Randall C.},
  title   = {Why there are complementary learning systems in the hippocampus and neocortex: insights from the successes and failures of connectionist models of learning and memory},
  journal = {Psychological Review},
  year    = {1995},
  volume  = {102},
  number  = {3},
  pages   = {419--457},
  doi     = {10.1037/0033-295X.102.3.419}
}

@article{kumaran2016complementary,
  author  = {Kumaran, Dharshan and Hassabis, Demis and McClelland, James L.},
  title   = {What Learning Systems do Intelligent Agents Need? Complementary Learning Systems Theory Updated},
  journal = {Trends in Cognitive Sciences},
  year    = {2016},
  volume  = {20},
  number  = {7},
  pages   = {512--534},
  doi     = {10.1016/j.tics.2016.05.004}
}

@article{wilson1994reactivation,
  author  = {Wilson, Matthew A. and McNaughton, Bruce L.},
  title   = {Reactivation of hippocampal ensemble memories during sleep},
  journal = {Science},
  year    = {1994},
  volume  = {265},
  number  = {5172},
  pages   = {676--679},
  doi     = {10.1126/science.8036517}
}

@article{yassa2011pattern,
  author  = {Yassa, Michael A. and Stark, Craig E. L.},
  title   = {Pattern separation in the hippocampus},
  journal = {Trends in Neurosciences},
  year    = {2011},
  volume  = {34},
  number  = {10},
  pages   = {515--525},
  doi     = {10.1016/j.tins.2011.06.006}
}

@article{diekelmann2010memory,
  author  = {Diekelmann, Susanne and Born, Jan},
  title   = {The memory function of sleep},
  journal = {Nature Reviews Neuroscience},
  year    = {2010},
  volume  = {11},
  number  = {2},
  pages   = {114--126},
  doi     = {10.1038/nrn2762}
}

@article{rao1999predictive,
  author  = {Rao, Rajesh P. N. and Ballard, Dana H.},
  title   = {Predictive coding in the visual cortex: a functional interpretation of some extra-classical receptive-field effects},
  journal = {Nature Neuroscience},
  year    = {1999},
  volume  = {2},
  number  = {1},
  pages   = {79--87},
  doi     = {10.1038/4580}
}

@article{srinivasan1982predictive,
  author  = {Srinivasan, Mandyam V. and Laughlin, Simon B. and Dubs, A.},
  title   = {Predictive coding: a fresh view of inhibition in the retina},
  journal = {Proceedings of the Royal Society of London. Series B},
  year    = {1982},
  volume  = {216},
  number  = {1205},
  pages   = {427--459},
  doi     = {10.1098/rspb.1982.0085}
}

@article{walsh2020evaluating,
  author  = {Walsh, Kevin S. and McGovern, David P. and Clark, Andy and O'Connell, Redmond G.},
  title   = {Evaluating the neurophysiological evidence for predictive processing as a model of perception},
  journal = {Annals of the New York Academy of Sciences},
  year    = {2020},
  volume  = {1464},
  number  = {1},
  pages   = {242--268},
  doi     = {10.1111/nyas.14321}
}

@article{hosoya2005dynamic,
  author  = {Hosoya, Toshihiko and Baccus, Stephen A. and Meister, Markus},
  title   = {Dynamic predictive coding by the retina},
  journal = {Nature},
  year    = {2005},
  volume  = {436},
  number  = {7047},
  pages   = {71--77},
  doi     = {10.1038/nature03689}
}

@article{parisi2019continual,
  author  = {Parisi, German I. and Kemker, Ronald and Part, Jose L. and Kanan, Christopher and Wermter, Stefan},
  title   = {Continual lifelong learning with neural networks: A review},
  journal = {Neural Networks},
  year    = {2019},
  volume  = {113},
  pages   = {54--71},
  doi     = {10.1016/j.neunet.2019.01.012}
}

@article{kirkpatrick2017overcoming,
  author  = {Kirkpatrick, James and Pascanu, Razvan and Rabinowitz, Neil and Veness, Joel and Desjardins, Guillaume and Rusu, Andrei A. and Milan, Kieran and Quan, John and Ramalho, Tiago and Grabska-Barwinska, Agnieszka and Hassabis, Demis and Clopath, Claudia and Kumaran, Dharshan and Hadsell, Raia},
  title   = {Overcoming catastrophic forgetting in neural networks},
  journal = {Proceedings of the National Academy of Sciences},
  year    = {2017},
  volume  = {114},
  number  = {13},
  pages   = {3521--3526},
  doi     = {10.1073/pnas.1611835114}
}

@inproceedings{ba2016fastweights,
  author    = {Ba, Jimmy and Hinton, Geoffrey E. and Mnih, Volodymyr and Leibo, Joel Z. and Ionescu, Catalin},
  title     = {Using Fast Weights to Attend to the Recent Past},
  booktitle = {Advances in Neural Information Processing Systems 29 (NIPS 2016)},
  year      = {2016},
  pages     = {4331--4339}
}

@article{johansson1979tactile,
  author  = {Johansson, R. S. and Vallbo, A. B.},
  title   = {Tactile sensibility in the human hand: relative and absolute densities of four types of mechanoreceptive units in glabrous skin},
  journal = {The Journal of Physiology},
  year    = {1979},
  volume  = {286},
  pages   = {283--300},
  doi     = {10.1113/jphysiol.1979.sp012619}
}

@article{luppi2024trainees,
  author  = {Luppi, Andrea I. and Achterberg, Jascha and Schmidgall, Samuel and Bilgin, Isil Poyraz and Herholz, Peer and Sprang, Maximilian and others},
  title   = {Trainees' perspectives and recommendations for catalyzing the next generation of {NeuroAI} researchers},
  journal = {Nature Communications},
  year    = {2024},
  volume  = {15},
  number  = {1},
  pages   = {9152},
  doi     = {10.1038/s41467-024-53375-2}
}

@article{ha2018worldmodels,
  author  = {Ha, David and Schmidhuber, J{\"u}rgen},
  title   = {World Models},
  journal = {arXiv preprint arXiv:1803.10122},
  year    = {2018},
  note    = {Also in Advances in Neural Information Processing Systems 31 (2018)}
}

@article{hafner2025dreamerv3,
  author  = {Hafner, Danijar and Pasukonis, Jurgis and Ba, Jimmy and Lillicrap, Timothy},
  title   = {Mastering diverse control tasks through world models},
  journal = {Nature},
  year    = {2025},
  volume  = {640},
  number  = {8059},
  pages   = {647--653},
  doi     = {10.1038/s41586-025-08744-2}
}

@article{schrittwieser2020muzero,
  author  = {Schrittwieser, Julian and Antonoglou, Ioannis and Hubert, Thomas and Simonyan, Karen and Sifre, Laurent and Schmitt, Simon and Guez, Arthur and Lockhart, Edward and Hassabis, Demis and Graepel, Thore and Lillicrap, Timothy and Silver, David},
  title   = {Mastering {Atari}, {Go}, chess and shogi by planning with a learned model},
  journal = {Nature},
  year    = {2020},
  volume  = {588},
  number  = {7839},
  pages   = {604--609},
  doi     = {10.1038/s41586-020-03051-4}
}

@article{miki2022locomotion,
  author  = {Miki, Takahiro and Lee, Joonho and Hwangbo, Jemin and Wellhausen, Lorenz and Koltun, Vladlen and Hutter, Marco},
  title   = {Learning robust perceptive locomotion for quadrupedal robots in the wild},
  journal = {Science Robotics},
  year    = {2022},
  volume  = {7},
  number  = {62},
  pages   = {eabk2822},
  doi     = {10.1126/scirobotics.abk2822}
}

@article{kaufmann2023drone,
  author  = {Kaufmann, Elia and Bauersfeld, Leonard and Loquercio, Antonio and M{\"u}ller, Matthias and Koltun, Vladlen and Scaramuzza, Davide},
  title   = {Champion-level drone racing using deep reinforcement learning},
  journal = {Nature},
  year    = {2023},
  volume  = {620},
  number  = {7976},
  pages   = {982--987},
  doi     = {10.1038/s41586-023-06419-4}
}

@article{silver2018alphazero,
  author  = {Silver, David and Hubert, Thomas and Schrittwieser, Julian and Antonoglou, Ioannis and Lai, Matthew and Guez, Arthur and Lanctot, Marc and Sifre, Laurent and Kumaran, Dharshan and Graepel, Thore and Lillicrap, Timothy and Simonyan, Karen and Hassabis, Demis},
  title   = {A general reinforcement learning algorithm that masters chess, shogi, and {Go} through self-play},
  journal = {Science},
  year    = {2018},
  volume  = {362},
  number  = {6419},
  pages   = {1140--1144},
  doi     = {10.1126/science.aar6404}
}

@article{mnih2015dqn,
  author  = {Mnih, Volodymyr and Kavukcuoglu, Koray and Silver, David and Rusu, Andrei A. and Veness, Joel and Bellemare, Marc G. and Graves, Alex and Riedmiller, Martin and Fidjeland, Andreas K. and Ostrovski, Georg and others},
  title   = {Human-level control through deep reinforcement learning},
  journal = {Nature},
  year    = {2015},
  volume  = {518},
  number  = {7540},
  pages   = {529--533},
  doi     = {10.1038/nature14236}
}

@article{izquierdo2013connectome,
  author  = {Izquierdo, Eduardo J. and Beer, Randall D.},
  title   = {Connecting a connectome to behavior: an ensemble of neuroanatomical models of {C. elegans} klinotaxis},
  journal = {PLoS Computational Biology},
  year    = {2013},
  volume  = {9},
  number  = {2},
  pages   = {e1002890},
  doi     = {10.1371/journal.pcbi.1002890}
}

@article{szigeti2014openworm,
  author  = {Szigeti, Bal{\'a}zs and Gleeson, Padraig and Vella, Michael and Khayrulin, Sergey and Palyanov, Andrey and Hokanson, Jim and Currie, Michael and Cantarelli, Matteo and Idili, Giovanni and Larson, Stephen},
  title   = {{OpenWorm}: an open-science approach to modeling {Caenorhabditis elegans}},
  journal = {Frontiers in Computational Neuroscience},
  year    = {2014},
  volume  = {8},
  pages   = {137},
  doi     = {10.3389/fncom.2014.00137}
}

@article{lappalainen2024connectome,
  author  = {Lappalainen, Janne K. and Tschopp, Fabian D. and Prakhya, Sridhama and McGill, Mason and Nern, Aljoscha and Shinomiya, Kazunori and Takemura, Shin-Ya and Gruntman, Eyal and Macke, Jakob H. and Turaga, Srinivas C.},
  title   = {Connectome-constrained networks predict neural activity across the fly visual system},
  journal = {Nature},
  year    = {2024},
  volume  = {634},
  number  = {8036},
  pages   = {1132--1140},
  doi     = {10.1038/s41586-024-07939-3}
}

@article{shiu2024flybrain,
  author  = {Shiu, Philip K. and Sterne, Gabriella R. and Spiller, Nico and Franconville, Romain and others},
  title   = {A {Drosophila} computational brain model reveals sensorimotor processing},
  journal = {Nature},
  year    = {2024},
  volume  = {634},
  number  = {8032},
  pages   = {210--219},
  doi     = {10.1038/s41586-024-07763-9}
}

@article{dorkenwald2024flywire,
  author  = {Dorkenwald, Sven and Matsliah, Arie and Sterling, Amy R. and Schlegel, Philipp and Yu, Szi-Chieh and McKellar, Claire E. and others},
  title   = {Neuronal wiring diagram of an adult brain},
  journal = {Nature},
  year    = {2024},
  volume  = {634},
  number  = {8032},
  pages   = {124--138},
  doi     = {10.1038/s41586-024-07558-y}
}

@article{scheffer2020hemibrain,
  author  = {Scheffer, Louis K. and Xu, C. Shan and Januszewski, Michal and Lu, Zhiyuan and Takemura, Shin-Ya and others},
  title   = {A connectome and analysis of the adult {Drosophila} central brain},
  journal = {eLife},
  year    = {2020},
  volume  = {9},
  pages   = {e57443},
  doi     = {10.7554/eLife.57443}
}

@article{Siegle2026LargeScale,
  author  = {Siegle, Joshua H. and Steinmetz, Nicholas A.},
  title   = {Large-scale electrophysiology at single-spike resolution},
  journal = {Nature Reviews Neuroscience},
  year    = {2026},
  volume  = {27},
  number  = {6},
  pages   = {395--415},
  doi     = {10.1038/s41583-026-01042-4}
}

@article{boyden2005optogenetics,
  author  = {Boyden, Edward S. and Zhang, Feng and Bamberg, Ernst and Nagel, Georg and Deisseroth, Karl},
  title   = {Millisecond-timescale, genetically targeted optical control of neural activity},
  journal = {Nature Neuroscience},
  year    = {2005},
  volume  = {8},
  number  = {9},
  pages   = {1263--1268},
  doi     = {10.1038/nn1525}
}

@article{brooks1986subsumption,
  author  = {Brooks, Rodney A.},
  title   = {A robust layered control system for a mobile robot},
  journal = {IEEE Journal of Robotics and Automation},
  year    = {1986},
  volume  = {2},
  number  = {1},
  pages   = {14--23},
  doi     = {10.1109/JRA.1986.1087032}
}

@article{lennie2003cost,
  author  = {Lennie, Peter},
  title   = {The cost of cortical computation},
  journal = {Current Biology},
  year    = {2003},
  volume  = {13},
  number  = {6},
  pages   = {493--497},
  doi     = {10.1016/S0960-9822(03)00135-0}
}

@article{attwell2001energybudget,
  author  = {Attwell, David and Laughlin, Simon B.},
  title   = {An energy budget for signaling in the grey matter of the brain},
  journal = {Journal of Cerebral Blood Flow and Metabolism},
  year    = {2001},
  volume  = {21},
  number  = {10},
  pages   = {1133--1145},
  doi     = {10.1097/00004647-200110000-00001}
}

@article{raymond1996cerebellum,
  author  = {Raymond, Jennifer L. and Lisberger, Stephen G. and Mauk, Michael D.},
  title   = {The cerebellum: a neuronal learning machine?},
  journal = {Science},
  year    = {1996},
  volume  = {272},
  number  = {5265},
  pages   = {1126--1131},
  doi     = {10.1126/science.272.5265.1126}
}

@article{HubelWiesel1962,
  author  = {Hubel, David H. and Wiesel, Torsten N.},
  title   = {Receptive Fields, Binocular Interaction and Functional Architecture in the Cat's Visual Cortex},
  journal = {The Journal of Physiology},
  year    = {1962},
  volume  = {160},
  number  = {1},
  pages   = {106--154},
  doi     = {10.1113/jphysiol.1962.sp006837}
}

@article{Fukushima1980,
  author  = {Fukushima, Kunihiko},
  title   = {Neocognitron: A Self-Organizing Neural Network Model for a Mechanism of Pattern Recognition Unaffected by Shift in Position},
  journal = {Biological Cybernetics},
  year    = {1980},
  volume  = {36},
  number  = {4},
  pages   = {193--202},
  doi     = {10.1007/BF00344251}
}

@article{LeCunEtAl1998,
  author  = {LeCun, Yann and Bottou, L{\'e}on and Bengio, Yoshua and Haffner, Patrick},
  title   = {Gradient-Based Learning Applied to Document Recognition},
  journal = {Proceedings of the IEEE},
  year    = {1998},
  volume  = {86},
  number  = {11},
  pages   = {2278--2324},
  doi     = {10.1109/5.726791}
}

@article{LeCunEtAl1989,
  author  = {LeCun, Yann and Boser, Bernhard and Denker, John S. and Henderson, Donnie and Howard, Richard E. and Hubbard, Wayne and Jackel, Lawrence D.},
  title   = {Backpropagation Applied to Handwritten Zip Code Recognition},
  journal = {Neural Computation},
  year    = {1989},
  volume  = {1},
  number  = {4},
  pages   = {541--551},
  doi     = {10.1162/neco.1989.1.4.541}
}

@article{MoranDesimone1985,
  author  = {Moran, Jeffrey and Desimone, Robert},
  title   = {Selective Attention Gates Visual Processing in the Extrastriate Cortex},
  journal = {Science},
  year    = {1985},
  volume  = {229},
  number  = {4715},
  pages   = {782--784},
  doi     = {10.1126/science.4023713}
}

@inproceedings{BahdanauChoBengio2015,
  author    = {Bahdanau, Dzmitry and Cho, Kyunghyun and Bengio, Yoshua},
  title     = {Neural Machine Translation by Jointly Learning to Align and Translate},
  booktitle = {International Conference on Learning Representations (ICLR)},
  year      = {2015}
}
\end{twocolumn}

\end{document}